\documentclass[twocolumn]{aastex61}
\usepackage{amsmath}

\usepackage{units}
\newcommand{\gcc}{\ensuremath{\mathrm{g\,cm^{-3}}}} 
\newcommand{\rhoth}{\ensuremath{\rho_{\mathrm{th}}}} 
\newcommand{\logT}{\ensuremath{\log(T/\mathrm{K})}} 
\newcommand{\logRho}{\ensuremath{\log(\rho/\gcc)}} 
\newcommand{\kB}{\ensuremath{k_\mathrm{B}}} 
\newcommand{\code}[1]{\texttt{#1}}
\newcommand{\mesa}{\code{MESA}}
\newcommand{\MESA}{\mesa}

\newcommand{\Ye}{\ensuremath{Y_{\mathrm{e}}}} 
\newcommand{\EF}{\ensuremath{E_\mathrm{F}}} 

\newlength{\apjcolwidth}
\newlength{\figwidth}
\begin{document}

\title{Electron Captures on $^{14}{\rm N}$ as a Trigger for Helium Shell Detonations}

\author[0000-0002-4791-6724]{Evan B. Bauer}
\affiliation{Department of Physics, University of California, Santa Barbara, CA 93106, USA}
\correspondingauthor{Evan B. Bauer}
\email{ebauer@ucsb.edu}

\author[0000-0002-4870-8855]{Josiah Schwab}
\altaffiliation{Hubble Fellow}
\affiliation{Department of Astronomy and Astrophysics, University of California, Santa Cruz, CA 95064, USA}

\author{Lars Bildsten}
\affiliation{Department of Physics, University of California, Santa Barbara, CA 93106, USA}
\affiliation{Kavli Institute for Theoretical Physics, University of California, Santa Barbara, CA 93106, USA}

\submitjournal{The Astrophysical Journal}
\received{May 15, 2017}
\revised{July 10, 2017}
\accepted{July 13, 2017}

\begin{abstract}
White dwarfs (WDs) that accrete helium at rates $\sim 10^{-8} \, M_\odot \, \rm yr^{-1}$, such as those in close binaries with sdB stars, can accumulate large ($\gtrsim 0.1 \, M_\odot$) helium envelopes which are likely to detonate.  
We perform binary stellar evolution calculations of sdB+WD binary systems with \mesa, incorporating the important reaction chain $^{14}{\rm N}(e^-, \nu) {^{14}{\rm C}} ( \alpha, \gamma) {^{18}{\rm O}}$ (NCO), including a recent measurement for the ${^{14}{\rm C}} ( \alpha, \gamma) {^{18}{\rm O}}$ rate.
In large accreted helium shells, the NCO reaction chain leads to ignitions at the dense base of the freshly accreted envelope, in contrast to $3\alpha$ ignitions which occur away from the base of the shell.
In addition, at these accretion rates, the shells accumulate on a timescale comparable to their thermal
time, leading to an enhanced sensitivity of the outcome on the accretion rate history. Hence, time dependent accretion rates from binary stellar evolution are necessary to determine the helium layer mass at ignition.
We model the observed sdB+WD system CD~$-30^\circ 11223$ and find that the inclusion of these effects predicts ignition of a $0.153 \, M_\odot$ helium shell, nearly a factor of two larger than previous predictions.
A shell with this mass will ignite dynamically, a necessary condition for a helium shell detonation.
\end{abstract}


\section{Introduction}

Many accreting white dwarfs (WDs) are discovered when a thermonuclear instability (i.e.\ nova)
occurs on their surface. These outcomes depend on the accreting fuel, the
accretion rate, $\dot M$, and the WD mass. 
The growing AM CVn class of binaries are WDs
accreting from a Roche lobe filling helium donor \citep{Nelemans04}.
No hydrogen is seen.
Recent observations are beginning to unveil one possible class of
progenitors for these systems, tight sdB+WD binaries ({$P_{\rm orb} <
100 \, \rm min$}) that should make contact within the sdB star's
helium burning lifetime \citep{Geier13,Kupfer17}.
Due to the large He shells that are likely to accumulate
prior to the onset of the
initial thermonuclear instability, such systems are
of interest as potential environments for helium detonations that
can lead to ``.Ia'' supernovae or even double detonation
supernovae \citep{Nomoto82,Woosley94,Bildsten07,Shen09,Shen14,Woosley11,Brooks15}.

When accreting from the outer, unburned layers of a He
burning star, the isotope $^{14}\rm N$ is present with a mass fraction 
set by the initial stellar metallicity, $X_{14}\approx 0.01 (Z/0.02)$.
When the accretor is a WD, this $^{14}{\rm N}$ is an important isotope, 
as it captures an electron when densities above $1.25 \times 10^6 \, \rm{g \, cm^{-3}}$ are reached.
The resulting $^{14}$C then undergoes the reaction 
$^{14}{\rm C}(\alpha, \gamma) {^{18}{\rm O}}$ that can trigger a thermonuclear
flash \citep{Hashimoto86}. This process, known as the NCO chain, 
requires the accumulation of a dense shell prior to the initiation of any
thermonuclear instability, and is the subject of our study. 

\cite{Hashimoto86} showed that this reaction chain can lead to an earlier ignition
than expected from the $3 \alpha$ reaction alone when accreting He onto a He WD, 
and \cite{Iben87} and \cite{Shen09} noted its potential importance for accretion onto C/O WDs.
\cite{Woosley94} included the NCO chain in their models of sub-Chandrasekhar helium
detonations.
\cite{Piersanti01} discussed the influence of NCO burning on the  location of the ignition point for
large, degenerate He envelopes on C/O WDs formed at constant ${\dot M \approx 10^{-8}\, M_\odot \,{\rm yr^{-1}}}$.
They concluded that NCO burning only marginally decreased accumulated He layer mass, and noted
that NCO burning did not lift degeneracy and prevent instability.
\cite{Woosley11} 
highlighted the role of the NCO chain in their survey of C/O WDs accreting He
at \mbox{$\dot M = (1-10) \times 10^{-8} \, M_\odot \, {\rm yr^{-1}}$},
finding that the electron captures modify the neutron excess of the burned material 
and reduce the density at which the thermonuclear runaway initiates.
This $\dot M$  and WD mass regime is coincident with that realized in
the sdB donor star scenario~\citep{Iben87,Brooks15} and so needs a thorough investigation. 

Our exploration of NCO ignitions in accreted He envelopes on C/O WDs using \MESA\ confirms the importance of the NCO chain for systems accreting at rates corresponding to sdB+WD scenarios.
Section~\ref{sec:NCOchain} describes the relevant reaction rates used as input for \MESA, relying on the recent work of \cite{Paxton15} and \cite{Schwab15} for the electron capture physics and \cite{Johnson09} for $\alpha$ captures on $^{14}{\rm C}$.
Section~\ref{sec:constmdot} shows \MESA\ results for models at constant~$\dot M$ to explore broad trends in the influence of the NCO chain.
Section~\ref{sec:binary} shows \MESA\ results that include binary evolution with resulting variable accretion rates.
These binary results are qualitatively different from what is found at constant~$\dot M$, demonstrating the importance of self-consistent evolution coupling detailed binary evolution and accretion histories to modeling of the accreting WD up to ignition of the He.
The system CD~$-30^\circ 11223$ \citep{Geier13} serves as a case study that naturally illustrates the importance of models including both NCO reactions and realistic binary accretion histories.

\section{The NCO Reaction Chain}
\label{sec:NCOchain}

Unless otherwise specified,
all modeling presented in this work relies on \MESA\ version r8118
with reaction networks including weak reactions
between $^{14} \rm N$ and $^{14} \rm C$ as well as $\alpha$-capture
onto $^{14} \rm C$. This section describes the details of the rates
for these reactions, which together make up the complete NCO chain.

\subsection{Weak Reactions for $^{14}{\rm N}$ and $^{14}{\rm C}$}

Tabulated rates for the electron-capture and beta-decay reactions
linking $^{14}{\rm N}$ and $^{14}{\rm C}$ are not included in \MESA\ 
version r8118.  In order to incorporate these important rates, we used
a modified version of \MESA's on-the-fly weak reaction rate
capabilities \footnote{Our inlists and patches for \MESA\ r8118 will
  be made available at \url{http://mesastar.org}.}
\citep{Paxton15, Schwab15, Paxton16, Schwab16}.
The rate
of interest is that of the ground state (${^{14}{\rm N}: J^\pi = 1^+}$)
to ground state (${^{14}{\rm C}: J^\pi = 0^+}$) transition.  This a has
$Q$-value of $\unit[0.1565]{MeV}$, corresponding to a threshold
density of ${\rho_{\mathrm{th}} = \unit[1.156\times10^6]{\gcc}}$ for
electron fraction ${\Ye = 0.5}$. The $ft$-value for $^{14} \rm C$
beta decay is
${\log(ft/\mathrm{s}) = 9.04}$ \citep{AjzenbergSelove91}, with the
$ft$-value for electron capture being a factor of 3 lower,
corresponding to the ratio of the spin degeneracies
${(2J_{\rm N} + 1)/(2J_{\rm C} + 1)}$.

The previous application of these capabilities \citep{Schwab15, MartinezRodriguez2016} focused on the
high-density regime where the degenerate electrons are
ultra-relativistic, and therefore we must make one slight modification
to the treatment included in \MESA\ version r8118.  The
equations implemented assume that $G$, the Coulomb barrier factor, can
be approximated as a constant and thus removed from the phase space
integral.  This is true in the ultra-relativistic regime, where
$G=\exp(\pi\alpha Z) \approx 1.2$ as well as in the non-relativistic
regime, where $G = 2\pi\alpha Z \approx 0.32$
\citep{Fuller80}. However, the density where the electron Fermi energy
$E_{\rm F} \approx m_\mathrm{e} c^2$ is
$\rho \approx \unit[2\times10^6]{\gcc}$ (for $\Ye = 0.5$),
near $\rho_{\mathrm{th}}$. We are not in either
limiting regime.
\cite{Fuller85} remove $G$ from the integral by replacing
it with a suitably defined average value, $\left<G\right>$.  These
values vary with density and temperature; however, we find the choice of a
single average value $\left<G_{\beta}\right> = 0.75$ (for beta decay)
and $\left<G_{\mathrm{ec}}\right> = 0.95$ (for electron capture)
reproduces the rates calculated without removing $G$ from the
integral to within $\approx 10\%$ over the regime of interest
($5 \le \logRho \le 7$ and $7 \le \logT \le 9$).

\begin{figure}
\begin{center}
\includegraphics[width=1.0\apjcolwidth]{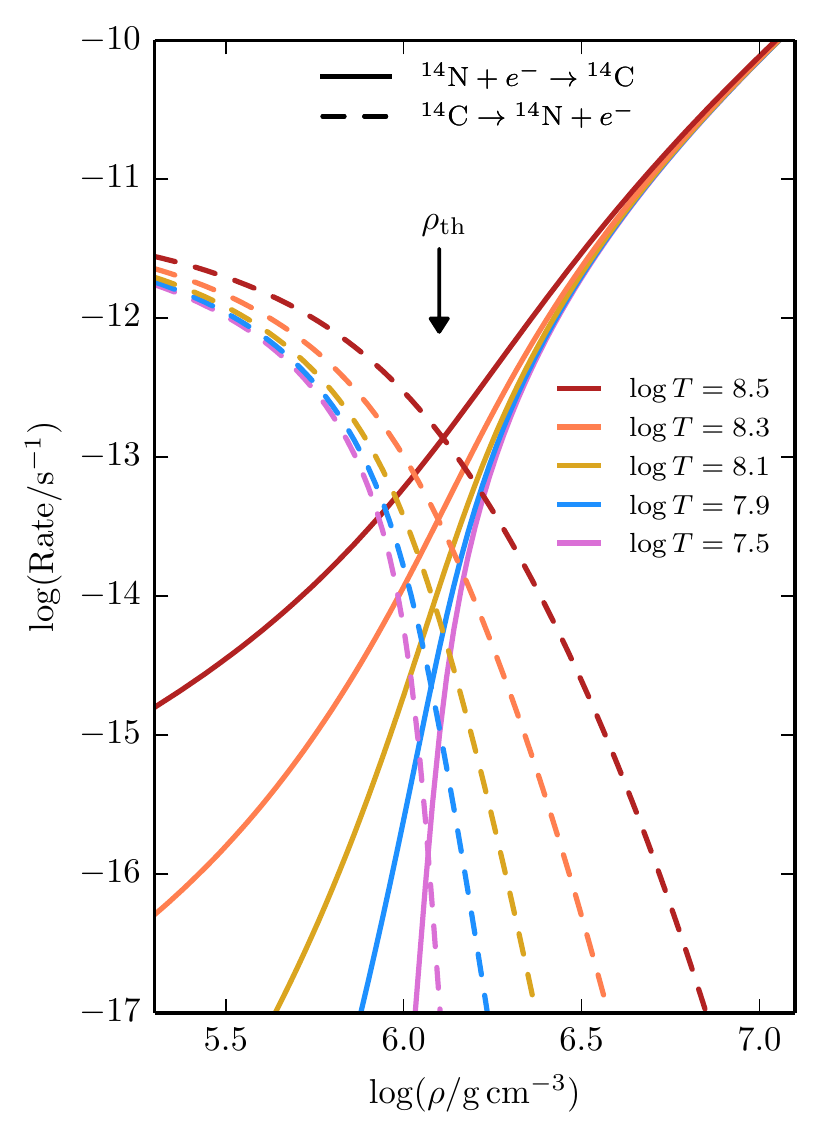}
\caption{
Rates for electron-capture and beta-decay reactions
linking $^{14} \rm N$ and $^{14} \rm C$ (for $\Ye = 0.5$).
So as to compare with previous work, this plot neglects the Coulomb correction.
}
\label{fig:elcap}
\end{center}
\end{figure}

Figure~\ref{fig:elcap} shows these weak reaction rates.  We confirmed
that these rates agree with the rates in \citet{Hashimoto86}
to within $\approx 10\%$ at the relevant, near-threshold densities.  The work of
\citet{Woosley11} used the \cite{Hashimoto86} results.  As discussed by \cite{Hashimoto86}, the most
important aspect of these rates is the shift in equilibrium
composition from $^{14}{\rm N}$ (at $\rho < \rhoth$) to $^{14}{\rm C}$
(at $\rho > \rhoth$) over a narrow range in density
$\Delta \log \rho \approx (\kB T)/\EF$.

The rates in our \MESA\ calculations include an additional
correction not present in the rates shown in Figure~\ref{fig:elcap};
this ``ion Coulomb correction'' corresponds to the energetic cost to
change the ion charge in the dense plasma.  We evaluate the magnitude
of this effect using the ion chemical potential from
\citet{Potekhin09a}.  At the densities and temperatures of interest,
this energy difference is
$\Delta \mu_{\mathrm{ion}} \approx \unit[7]{keV}$.  This corresponds
to an increase of the threshold density by
$\Delta \rhoth \approx \unit[10^5]{\gcc}$.  When referring to the
threshold density for our \MESA\ models, we use the value
$\rhoth = \unit[1.25 \times 10^6]{\gcc}$ which accounts for this
correction.

\subsection{The $^{14}{\rm C}(\alpha, \gamma) {^{18}{\rm O}}$ Rate}
\label{sec:c14_rate}

\begin{figure}
\begin{center}
\includegraphics[width=\apjcolwidth]{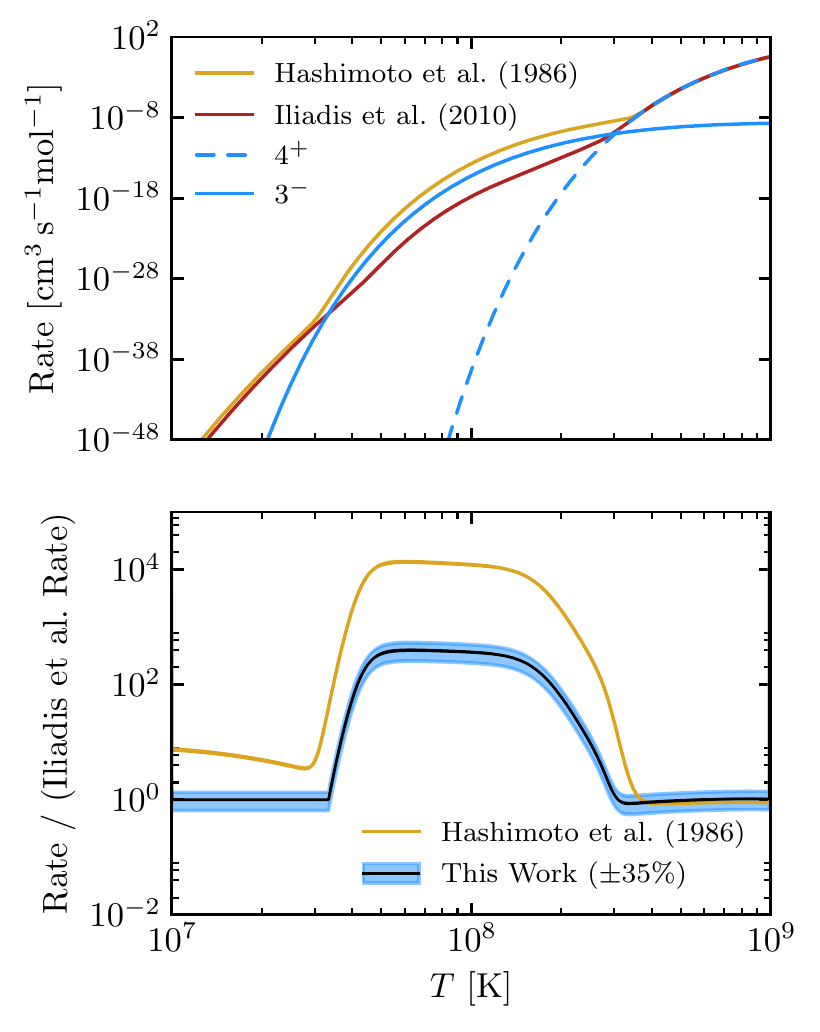}
\caption{
(Top)
The $^{14}{\rm C}(\alpha, \gamma) {^{18}{\rm O}}$ rates from
\cite{Hashimoto86} and \cite{Iliadis10}, along
with the specific resonances that dominate for $T> 3 \times 10^7 \, \rm K$
as measured by \cite{Johnson09}.
(Bottom) 
The $^{14}{\rm C}(\alpha, \gamma) {^{18}{\rm O}}$ rates from this work
and \cite{Hashimoto86} relative to the rate from \cite{Iliadis10}.
}
\label{fig:c14_rates}
\end{center}
\end{figure}

Historically, the $^{14}{\rm C}(\alpha, \gamma) {^{18}{\rm O}}$ rate
has been uncertain by several orders of magnitude due to a poorly
constrained, near-threshold, $3^-$ resonance in $^{18}{\rm O}$ at
$6.404 \,\rm MeV$, which dominates the rate for temperatures
${3 \times 10^7 \, {\rm K} < T < 3 \times 10^8 \, \rm K}$.
Figure~\ref{fig:c14_rates} shows the rate given in equation~(1) of
\cite{Hashimoto86}, as well as the rate from \cite{Iliadis10},
via the JINA Reaclib database \citep{Cyburt10}, that was
adopted as the default rate in \mesa\ \citep{Paxton11,
  Paxton13}. The contrast between these rates illustrates the large
historical uncertainty associated with the temperature regime dominated by the
resonance. 

For this work we use the measurements of \cite{Johnson09} for the
temperature regime $T > 3 \times 10^7 \, \rm K$, where the rate is
dominated by the $3^-$ and $4^+$
resonances. We have adopted the rates given in their equation~(12) for
those resonances, with a claimed
uncertainty of just $35\%$ for the $3^-$ resonance. Thus, the
historical uncertainty associated with the $^{14}{\rm
  C}(\alpha, \gamma) {^{18}{\rm O}}$ rate is now greatly reduced 
in the temperature regime relevant for our problem. The contributions
from these resonances are plotted in Figure~\ref{fig:c14_rates}
for comparison to the other full rates. For lower temperatures where
these resonances do not dominate, we switch back to using the rate
from \cite{Iliadis10} for simplicity, though the rate is so small in
this region that it will not be significant.
The lower panel of Figure~\ref{fig:c14_rates} shows the total resulting
rate that we have adopted for this work relative to the default
rate found in \mesa\ version r8118 from \cite{Iliadis10}.

\subsection{Example of He Accretion onto a He WD}

\begin{figure}
\begin{center}
\includegraphics[width=\apjcolwidth]{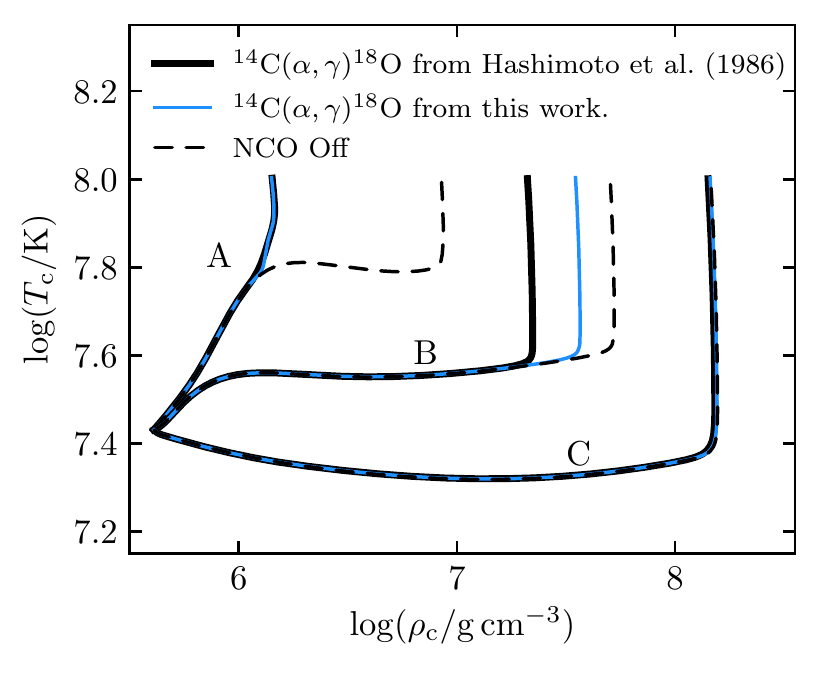}
\caption{
\MESA\ models of He accretion onto a He~WD
reproducing those shown by figure~5 in \cite{Hashimoto86}.
Solid black lines show core ignition via the NCO chain using
the $^{14}{\rm C}(\alpha, \gamma) {^{18}{\rm O}}$ rate given by
\cite{Hashimoto86} equation~(1). Dashed lines show ignition when NCO
burning is ignored and only $3\alpha$ plays a role. Solid blue lines
show the result from NCO ignition using the $^{14}{\rm C}(\alpha,
\gamma) {^{18}{\rm O}}$ rate described in Section~\ref{sec:c14_rate}
of this work. The three different cases are constant He accretion
rates of
(A) $10^{-8} \, M_\odot \, \rm yr^{-1}$,
(B) $10^{-9} \, M_\odot \, \rm yr^{-1}$, and
(C) $3 \times 10^{-10} \, M_\odot \, \rm yr^{-1}$.
}
\label{fig:hashimoto}
\end{center}
\end{figure}

To exhibit how \mesa\ compares to prior work,
we used \mesa\ to reproduce the He WD
evolution scenarios described in section~4 and figure~5 of
\cite{Hashimoto86}. A $0.3 \, M_\odot$ He WD model
accretes He until the center is compressed and heated enough to
undergo an NCO induced thermonuclear runaway.
In Figure~\ref{fig:hashimoto}, we compare cases with different rates
for the $^{14}{\rm C}(\alpha, \gamma) {^{18}{\rm O}}$ step in the NCO
chain, as well as a case where NCO reactions are omitted from the
network.
For higher accretion rates, the temperature of the core is high enough
that electron captures are the rate limiting step for the NCO chain,
and hence we see no difference in the evolution tracks
when using different $^{14}{\rm C}(\alpha, \gamma) {^{18}{\rm O}}$
rates. For lower accretion rates, however, the core evolution tracks
reach well beyond the threshold density for electron captures, so that
electron captures are no longer the rate limiting step for the NCO chain. Instead,
the tracks lie in a temperature region where $^{14} \rm C$ burning
dominates the net NCO rate, and we see that the improved
$^{14}{\rm C}(\alpha, \gamma) {^{18}{\rm O}}$ rate
\citep{Johnson09} substantially changes the final outcome for
case~B. In case~C, He burning triggers the thermonuclear runaway
before NCO has a chance, so the $^{14}{\rm C}(\alpha, \gamma)
{^{18}{\rm O}}$ rate ends up being irrelevant for igniting the flash.

\section{NCO Reactions and Helium Accretion}
\label{sec:constmdot}

Due to the steep density dependence of the electron capture rates, we expect the NCO chain to play a significant role only when the density at the base of an accreted He shell reaches values above the threshold density of $\rho_{\rm th} = 1.25 \times 10^6 \, \rm g \, cm^{-3}$ prior to thermonuclear ignition.
The rate at which NCO burning occurs can be governed by the electron captures
on $^{14}{\rm N}$ (and hence the local density), but most of the energy production from the chain is
supplied by the subsequent burning of $^{14}{\rm C}$.
Once the right conditions are reached for electron captures onto $^{14}{\rm N}$,
alpha captures occur on the freshly produced $^{14}\rm C$, releasing 
$Q = 6.227 \, {\rm MeV}$ per $^{14} \rm C$ consumed. At constant pressure and for ions strongly in the liquid state, complete 
consumption of the $^{14}\rm C$ at abundance $X_{14} \ll 1$ in a helium background leads to a temperature change of
\begin{equation} 
\label{eq:total_heat}
\Delta T =  \frac{2}{21} \frac{Q}{k_{\rm B}} X_{14} \approx 7 \times 10^7 \, {\rm K} \left( \frac{X_{14}}{0.01} \right). 
\end{equation} 
This entropy input is often large enough to trigger a full He burning runaway, and $3\alpha$ burning quickly takes
over as the dominant energy source once NCO has raised the temperature enough to initiate a runaway.

In contrast, when NCO reactions are ignored and ignition depends on $3\alpha$ reactions alone,
the models experience a later ignition in a different location. Helium burning via $3 \alpha$ is much more
temperature sensitive than the electron captures that initiate the NCO chain, which depend primarily on the density.
Despite previous work on mixing and viscous heating due to shear instabilities
for white dwarfs accreting helium \citep{Yoon04a,Yoon04c,Yoon04b},
we ignore these effects in our models. Recent work by \cite{Piro15} suggests that
this should be justified due to the baroclinic instability inhibiting
development of shear instabilities at depths relevant for helium ignition.

\subsection{Constant $\dot M$ Without NCO}

\begin{figure}
\begin{center}
\includegraphics[width=\apjcolwidth]{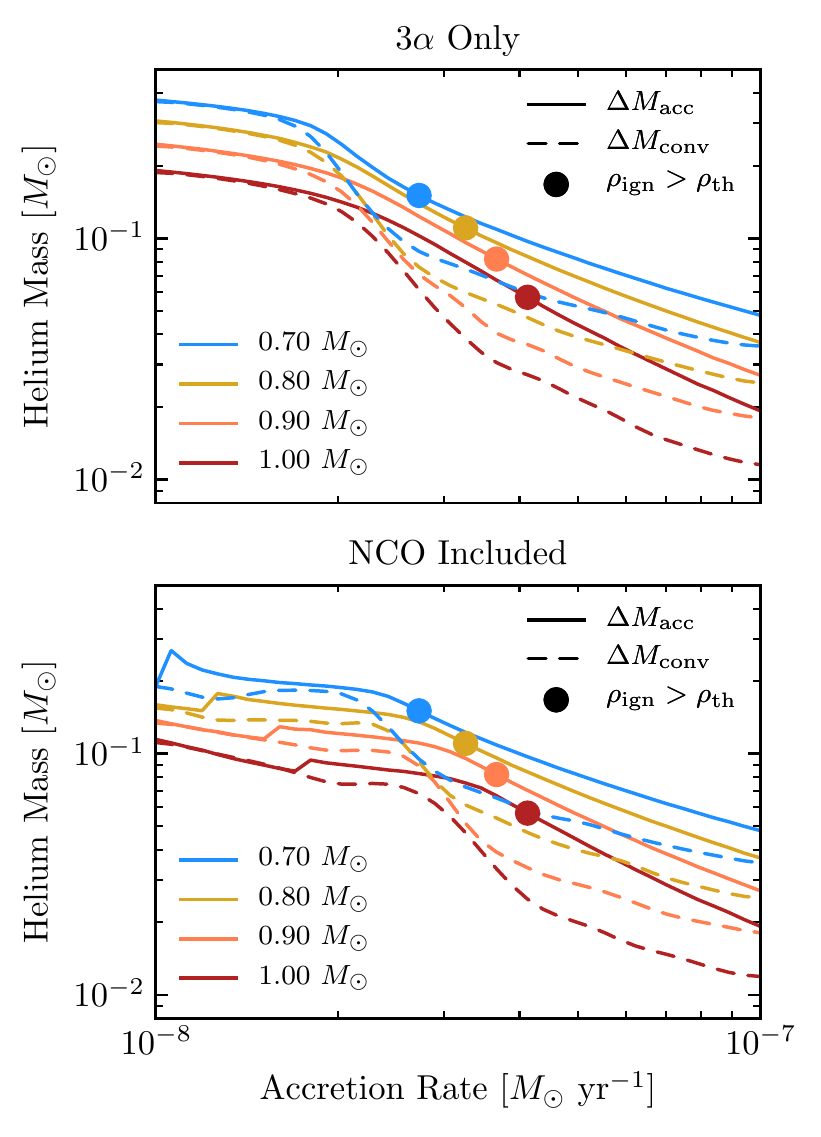}
\caption{
Accumulated helium shell masses and convective shell masses for
flashes on various WD masses over a range of constant accretion rates.
The upper panel shows flashes ignited by $3 \alpha$ alone,
while the lower panel shows flashes when NCO reactions are included.
Points indicate the first flash where the density at the
base of the accreted material
was above ${\rho_{\rm th} = 1.25\times 10^6 \,{\rm g \, cm^{-3}}}$.
}
\label{fig:he_shells}
\end{center}
\end{figure}

As a baseline for comparison, we first created a grid of WD models over a range of constant He accretion rates until they reached $3 \alpha$ ignition in their shells. 
During the accretion phase, the location of peak temperature lies outside the base of the accreted envelope due to the generic feature of
a temperature inversion at these accretion rates, where electron conduction competes with the compressional heating
by draining heat into the core and cooling the most dense inner layers of the envelope \citep{Nomoto82}.
Since temperature inversions can cause a $3 \alpha$ based runaway to happen at a location in
the accumulated material other than the base, the convective shell mass can be less than the total accreted mass.
The top panel of Figure~\ref{fig:he_shells} shows the mass of the accumulated (solid) and convective (dashed) He shells
at the moment of the instability triggered by the $3\alpha$ reactions alone. 
The convective shell mass is defined here as all mass exterior to the location of the thermonuclear runaway, which will
be swept up in the convection that occurs as a result of unstable ignition.

Models to the left of the solid point in Figure~\ref{fig:he_shells} accumulated sufficiently large He envelopes to achieve a density above
${1.25 \times 10^6 \, \rm{g \, cm^{-3}}}$ at the base of the He layer prior to $3 \alpha$ ignition, 
implying that for ${\dot M<4\times 10^{-8} \, M_\odot {\rm\,yr^{-1}}}$, 
the NCO reaction chain should provide extra heat near the dense base of the envelope.
For all models accreting at constant $\dot M$, we assume an initial core temperature of ${T_{\rm c} = 2 \times 10^7 \, \rm K}$,
appropriate for sdB+WD binary scenarios where the WD certainly has ${10 - 100 \, \rm Myr}$ to cool before the system makes contact. Lower core
temperatures do not significantly impact the results.

\subsection{Constant $\dot M$ With NCO}
\label{sec:constNCO}

A set of models similar to those shown in the top panel of Figure~\ref{fig:he_shells}, but now including NCO reactions, is shown in the lower panel of Figure~\ref{fig:he_shells}. For the region where density is beyond the threshold for NCO, we can see that the total accreted mass is somewhat lower due to earlier ignition, but in some cases the convective shell can still encompass more total mass due to the ignition occurring deeper in the accreted material.

\begin{figure}
\begin{center}
\includegraphics[width=\apjcolwidth]{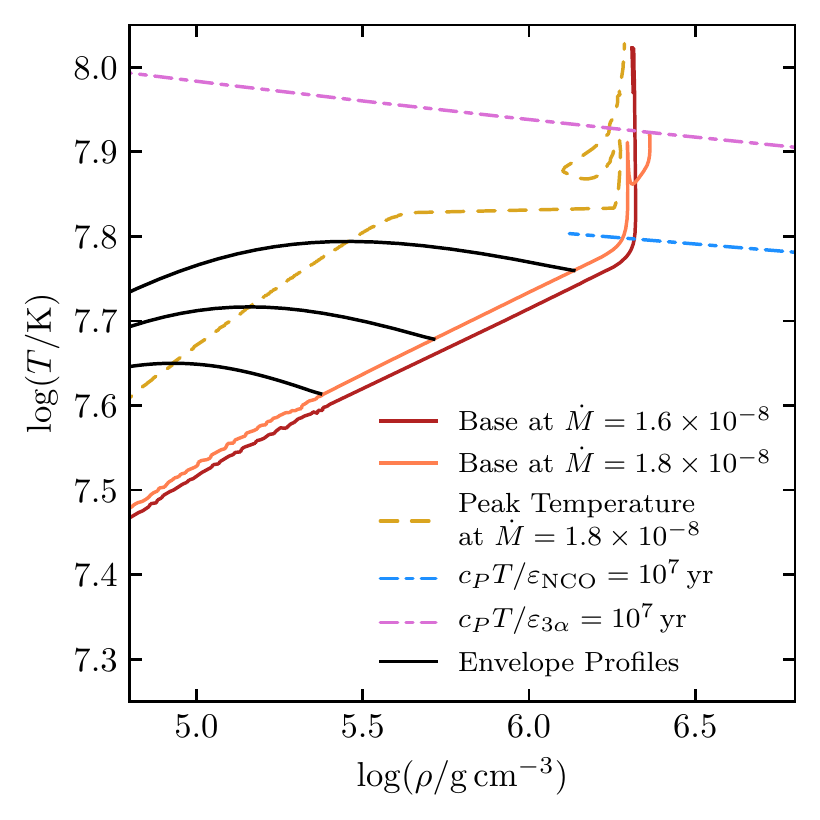}
\caption{
Density-temperature evolution of the accreted layer for a $1.0 \, M_\odot$ WD accreting at two different rates. These two models correspond to the discontinuity in the red line in the lower panel of Figure~\ref{fig:he_shells}.
}
\label{fig:failed}
\end{center}
\end{figure}

The discontinuous feature in total accreted mass at low $\dot M$ in the lower panel of Figure~\ref{fig:he_shells} is due to failed NCO ignitions for certain accretion rates. The finite supply of $^{14}{\rm C}$ at $X_{14} \approx 0.01$ can be exhausted before NCO burning can fully ignite a $3\alpha$ runaway. Due to the highly degenerate conditions at the base of He envelopes that are dense enough for NCO to occur, electron conduction can carry significant amounts of heat inward toward the cooler core as $^{14}{\rm C}(\alpha, \gamma) {^{18} \rm O}$ begins to run away. This leads to a $\Delta T$ smaller than that predicted by equation~(\ref{eq:total_heat}). Figure~\ref{fig:failed} shows the envelope $\rho-T$ evolution of two models at very similar $\dot M$, where one experiences a failed NCO runaway at its base before eventually experiencing a true $3 \alpha$ runaway at the peak temperature location further out in the accreted He envelope.

\begin{figure}
\begin{center}
\includegraphics[width=\apjcolwidth]{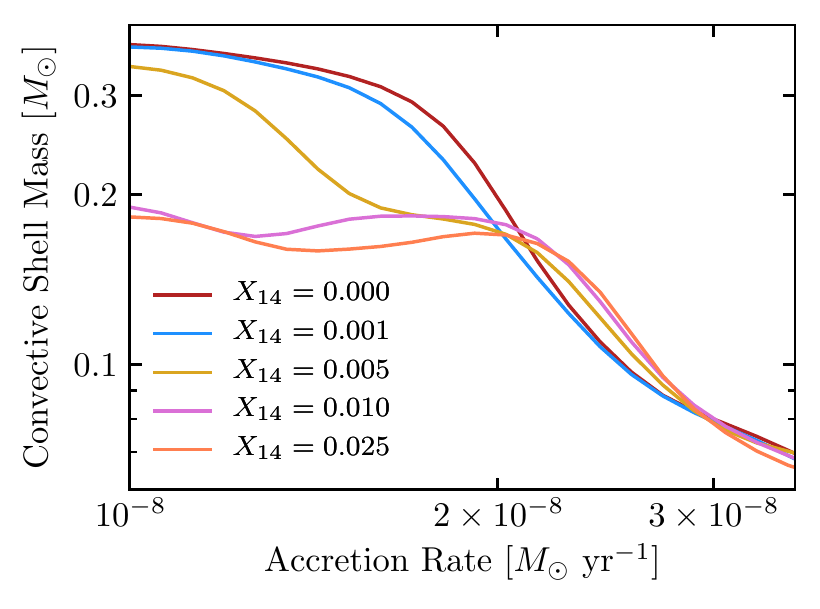}
\caption{
Convective shell masses at ignition as a function of constant accretion rate onto a $0.70 \, M_\odot$ WD for varying metallicity.
The lines shown here for ${X_{14} = 0.000}$ and ${X_{14} = 0.010}$ correspond to the dashed blue lines in Figure~\ref{fig:he_shells}.
}
\label{fig:metallicity}
\end{center}
\end{figure}

The metallicity sets the total amount of $^{14} \rm N$ available for NCO reactions. Since $\Delta T$ from complete NCO consumption scales with $X_{14}$ in equation~(\ref{eq:total_heat}), variation in metallicity directly corresponds to variation in the total thermal impact that NCO reactions can have.
Varying only the initial $^{14} \rm N$ content in the grid of constant $\dot M$ models reveals a strong dependence on metallicity.
These results are shown in Figure~\ref{fig:metallicity}.
For simplicity, we only present the variation in runs for the $0.7 \, M_\odot$ WD accretor model. Results for other accretor masses are similar.
We assume that both the donor and accretor were born with the same metallicity, and that all CNO elements from the initial metallicity eventually end up as $^{14}\rm N$ in both stars due to CNO burning in the evolution that produces them. Hence the initial mass fraction of $^{14} \rm N$ in the He envelope of the WD is correlated with that in the He accreted from the donor. Since the initial He envelope ends up as the base of the He layer after accretion, it contributes to the energy produced in the dense layers where NCO reactions occur.

Larger convective shell masses are associated with denser ignition locations that may be expected to experience dynamical burning. This provides the potential for developing detonation fronts that can give rise to interesting phenomenology such as .Ia supernovae \citep{Bildsten07,Shen09} or double detonation type Ia supernovae \citep{Woosley94} if the detonation can transition into the C/O core. The threshold envelope mass for dynamical burning is on the order of ${\sim 0.1 \, M_\odot}$.
Because the NCO chain increases the convective shell mass in some regions, and the lower accretion rate regime is associated with large total accumulation masses, we see that NCO reactions are important for systems that have the potential to ignite dynamically.

Realistic binary systems often have accretion rates  
that vary across the boundary for high-density ignition shown by solid dots in Figure~\ref{fig:he_shells}.
Furthermore, Figure~\ref{fig:he_shells} shows that small variations in accretion rate around ${2-4 \times 10^{-8} \, M_\odot \, \rm yr^{-1}}$
can significantly impact the final convective shell mass, determining the dynamical fate of ignition.
Clearly the constant $\dot M$ approximation is a concern, so we now
use the robust binary capabilities present in \MESA\ to test a range of realistic parameters and scenarios for NCO ignitions.

\newpage
\section{Realistic Mass Transfer Scenarios}
\label{sec:binary}

We now show that NCO burning plays a significant role in the initial
flash encountered in He star or sdB donor systems when the WD builds up a large
He envelope.
Our results at constant $\dot M$ in Section~\ref{sec:constNCO} suggest that the main impact
of NCO reactions is to decrease both total accreted mass and convective envelope mass
at low $\dot M$.
However, our  simulations of realistic binary evolution scenarios indicate
that NCO burning can be much more significant than the constant $\dot M$ results suggest.
\cite{Neunteufel16} studied systems like these using
detailed binary evolution and accretion rates while drawing on
\cite{Woosley11} for ignition outcomes of WDs treated as point mass accretors.
However, our results indicate that the constant
$\dot M$ results have limited predictive power in binary systems. Both the system
and the WD must be evolved.

\subsection{The First Flash after Contact}
\label{sec:FirstFlash}

\cite{Brooks15} used \MESA's binary evolution capabilities to model AM CVn systems,
including realistic accretion histories for systems that are brought into contact by
gravitational wave radiation,
with self-consistent binary stellar evolution tracked through the accretion phase.
Their study included many cycles of accretion, ignition, and flashes, but NCO reactions were
not included. For many of the flashes, they found accumulated masses that were
insufficient for the He layer to reach densities required for NCO reactions. A few of the flashes,
however, did accumulate sufficient mass, particularly those occuring after the system first comes
into contact and has not yet been warmed by previous flash episodes. In this section, we re-examine two
of these binary scenarios where NCO reactions can play a role. The first is a $0.4 \, M_\odot$ He star
donating onto a $0.8 \, M_\odot$ WD, and the second is the same donor model with a $1.0 \, M_\odot$ WD
accretor. These correspond to panels 2 and 4 in figures 12 and 13 from \cite{Brooks15}.

The accretion rate is primarily governed by the physics of the donor star, and our study here
leaves this unmodified, so we use the same $\dot M$ histories as presented
in figure~12 of \cite{Brooks15}
up to the point of ignition. With an identical starting model for the WD accretor and the accretion rate as
specified by previous \mesa\ binary runs,
it is sufficient to follow the single star evolution in \mesa\ for the accretor, with no further need to
invoke \mesa\ binary.

\begin{figure}
\begin{center}
\includegraphics[width=\apjcolwidth]{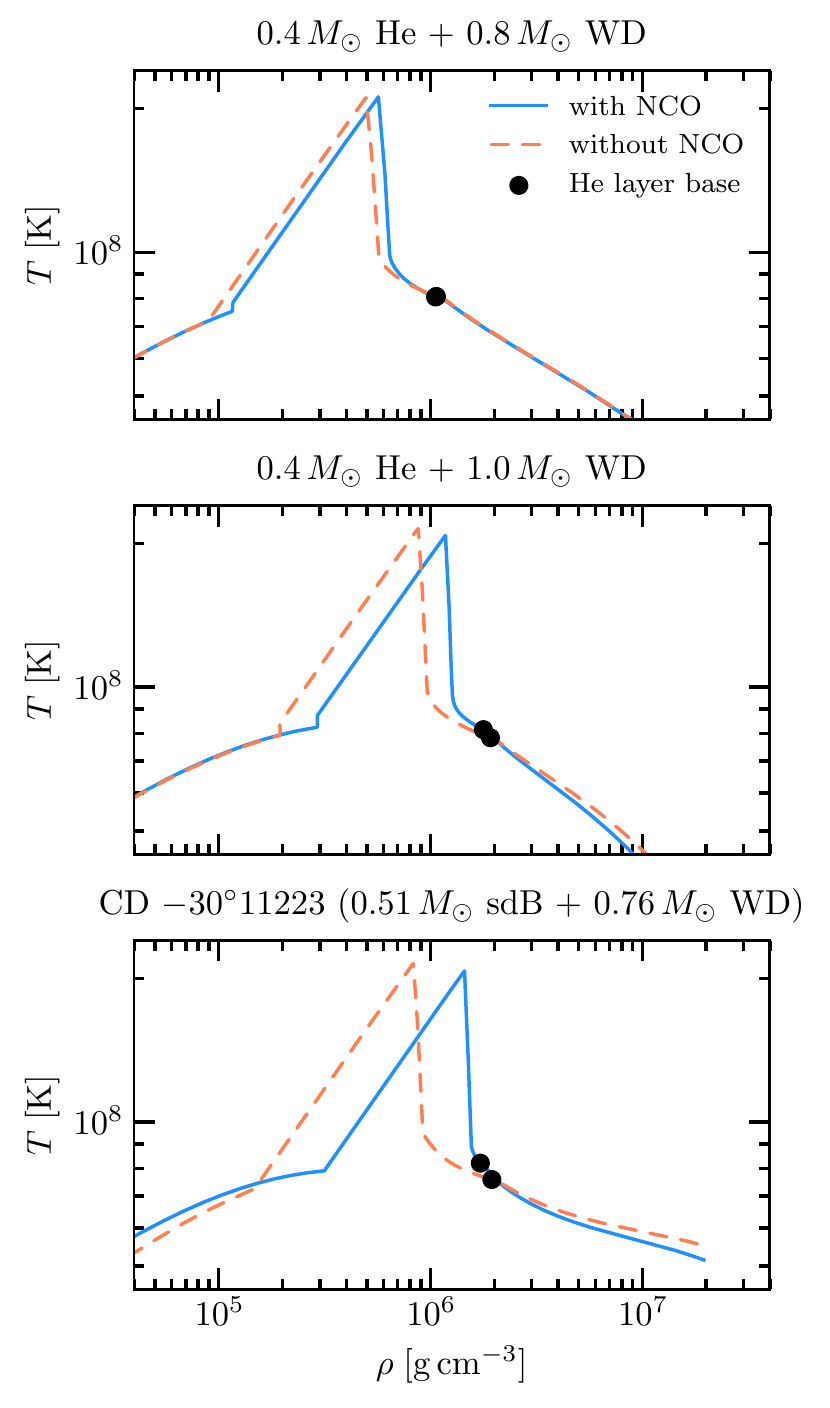}
\caption{
Profiles from models of the accreting WD
in several binary systems
after the flash
has ignited. Models that include NCO reactions ignite in
the deeper, denser region.
In the case of CD~$-30^\circ 11223$, the outer envelope has been
noticeably warmed by additional heat from NCO burning
prior to thermonuclear runaway.
}
\label{fig:CD30ign}
\end{center}
\end{figure}

Figure~\ref{fig:CD30ign} and Table~\ref{tab:mass} show that NCO reactions can modify the thermal structure prior
to ignition, and more importantly, lead to ignition in the much deeper layers near the base of the accreted He.
This effect is more pronounced for some systems than others, and the accretion rate from binary evolution
plays a large role in determining the thermal structure of the accreting WD, which governs the impact
of NCO reactions. The accretion rate varies from ${2-4 \times 10^{-8} \, M_\odot \, \rm yr^{-1}}$ over the course of accumulation
leading to the first flash (see e.g.\ figures 11 and 12 in \citealp{Brooks15}).
Since this $\dot M$ range is precisely where Figure~\ref{fig:he_shells} shows the most significant
variation in convective shell mass, there appears to be no
reliable way to estimate the impact of NCO on ignition based on results at constant $\dot M$.
Thus, accretion histories from full, self-consistent binary
evolution such as those provided here need to be
used to assess the condition at the time of thermal runaway.

\begin{table}
\caption{Accreted and convective masses for the first helium flash}
\label{tab:mass}
\begin{center}
\begin{tabular}{ | c | c  c | }
\hline
Accreted Mass ($\Delta M_{\rm acc}$)  & Without NCO & With NCO \\
\hline 
$0.4 \, M_\odot$ He + $0.8 \, M_\odot$ WD & $0.107 \, M_\odot$ & $0.107 \, M_\odot$ \\
\hline
$0.4 \, M_\odot$ He + $1.0 \, M_\odot$ WD & $0.082 \, M_\odot$ & $0.078 \, M_\odot$ \\
\hline
CD~$-30^\circ 11223$ & $0.175 \, M_\odot$ & $0.163 \, M_\odot$ \\
\hline \hline
Convective Mass ($\Delta M_{\rm conv}$) & Without NCO & With NCO \\
\hline 
$0.4 \, M_\odot$ He + $0.8 \, M_\odot$ WD & $0.055 \, M_\odot$ & $0.064 \, M_\odot$ \\
\hline
$0.4 \, M_\odot$ He + $1.0 \, M_\odot$ WD & $0.039 \, M_\odot$ & $0.054 \, M_\odot$ \\
\hline
CD~$-30^\circ 11223$ & $0.084 \, M_\odot$ & $0.153 \, M_\odot$ \\
\hline
\end{tabular}
\end{center}
\end{table}

\subsection{The First Flash in CD~$-30^\circ 11223$}
\label{sec:CD-30}

CD~$-30^\circ 11223$ \citep{Geier13} is a sdB+WD binary
system with an orbital period of 70.5 minutes
that will make contact in $40$~Myr, likely while the sdB star's core is
still burning He. \cite{Brooks15} used \mesa\ to model
the binary evolution of this system as
a $0.510 \, M_\odot$ sdB star donating onto a $0.762 \, M_\odot$ WD,
with initial conditions tuned to match the observations of \cite{Geier13}
assuming that the sdB star is just beginning helium core burning.
They predicted that the WD would
accumulate a large He  envelope (${\Delta M_{\rm acc} \approx 0.175 \, M_\odot}$)
which will then experience $3\alpha$ ignition above the 
base of envelope, leading to a smaller convective envelope
(${\Delta M_{\rm conv} \approx 0.084 \, M_\odot}$).
For more information on the details of the modeling of this binary system, see
section 3.2 in \cite{Brooks15}.
Using the same accretion history and starting model, we modeled the evolution
of the accreting WD both with and without NCO reactions.
If we do not include the reactions necessary for the NCO chain in our network,
our results match those described by \cite{Brooks15}.
In contrast, with NCO burning included in the network, 
the extra heat injected in the deeper, denser layers of
the envelope leads to an earlier ignition of a slightly smaller
(${\Delta M_{\rm acc} = 0.163 \, M_{\odot}}$) He envelope triggered by
$^{14}{\rm C}$.
However, since the ignition is triggered much deeper
in the accreted envelope, as seen in Figure~\ref{fig:CD30ign},
this results in a much larger, and more dynamically important, 
convective envelope of mass ${\Delta M_{\rm conv} = 0.153 \, M_\odot}$.

This near doubling of the convective shell mass has no parallel from
the results at constant $\dot M$ shown in Section~\ref{sec:constNCO}. In fact the trend seen there is in the opposite direction,
where Figure~\ref{fig:he_shells} shows that NCO ignitions mostly tend to suppress the size of the
convective shell by causing an earlier ignition while less total helium has had a chance to accumulate.
This qualitatively different result of a much larger convective shell further motivates the use of full binary
calculations in \mesa\ to avoid the approximation of constant $\dot M$.

\begin{figure}
\begin{center}
\includegraphics[width=\apjcolwidth]{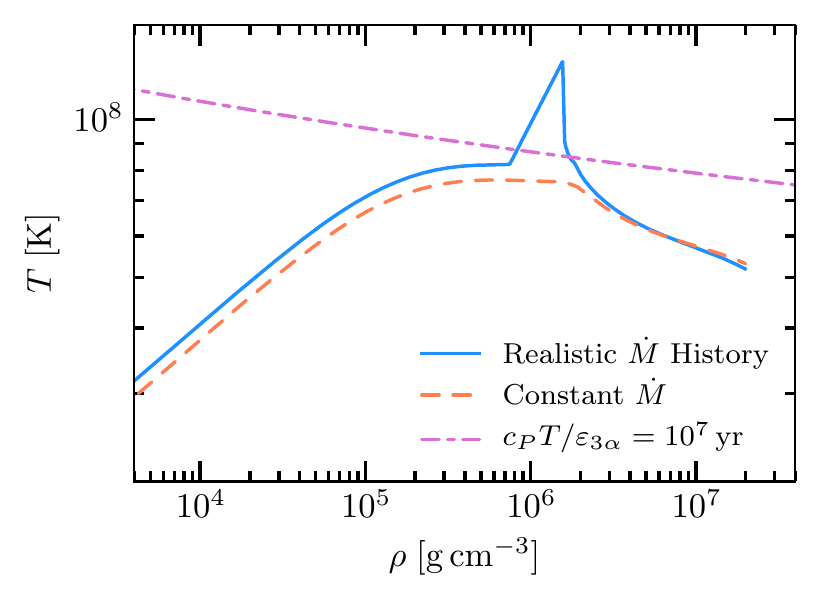}
\caption{
Effect of varying accretion history for CD~$-30^\circ 11223$.  The solid
density-temperature profile shows the model evolved with a realistic
accretion history from binary evolution.  The dashed profile shows a
model with a constant $\dot M$, corresponding to the time-average of the
realistic history.  The profiles are shown when the models reach a total
mass of ${M = 0.925 \, \rm M_\odot}$.
}
\label{fig:accretion}
\end{center}
\end{figure}

Figure~\ref{fig:accretion} shows the contrast between modeling of CD~$-30^\circ 11223$
including realistic accretion histories and modeling that makes the approximation of
constant $\dot M$. The latter case assumes ${\dot M = 1.93 \times 10^{-8} \, M_\odot \, \rm yr^{-1}}$,
the time-average of the accretion rate from the binary evolution calculations of \cite{Brooks15}.
Both models here include the NCO reaction chain.
After accumulating the same amount of mass to reach ${M = 0.925 \, M_\odot}$, the model with
constant $\dot M$ has not yet reached $3\alpha$ ignition, while the realistic $\dot M$ model has.
Indeed, the constant $\dot M$ model must accumulate $3 \%$ more mass
before reaching ignition at a final mass of ${M = 0.930 \, M_\odot}$.

\subsection{Metallicity and $^{14} \rm N$ Abundance in sdB Donors}

For sdB star donors, the interior abundance of $^{14} \rm N$ may be somewhat lower due to burning during the helium core flash having processed some of the $^{14} \rm N$ to $^{18} \rm O$ and $^{22} \rm Ne$.
\mesa\ models of the helium core flash show about half of the $^{14} \rm N$ in the interior of the He core will be consumed, leaving behind $X_{14} \approx 0.005$ of the original $X_{14} \approx 0.01$ for resulting sdB models (at solar metallicity). The abundance remains $X_{14} \approx \rm 0.01$ only in the unprocessed outer $\sim 0.01 \, M_\odot$; see figure~42 of \cite{Paxton15}.
This means that after the first $0.01 \, M_\odot$ of material donated, the abundance of $^{14} \rm N$ in the donated material will drop to $X_{14} \approx 0.005$.
However, NCO reactions are primarily significant in material from the initial He envelope of the accretor and first donated material, which will eventually form the most dense region of the He layer at its base. As Table~\ref{tab:mass} shows, these NCO reactions at the base of the He layer can provide sufficient heating to ignite the envelope within $\approx 0.01 \, M_\odot$ of its base, so the lower abundance of $^{14} \rm N$ in the accreted material further out should not significantly impact the outcome, and we ignore this change in abundance for all models presented in this work.

\section{Dynamical Burning in Large Helium Envelopes}

\begin{figure}
\begin{center}
\includegraphics[width=\apjcolwidth]{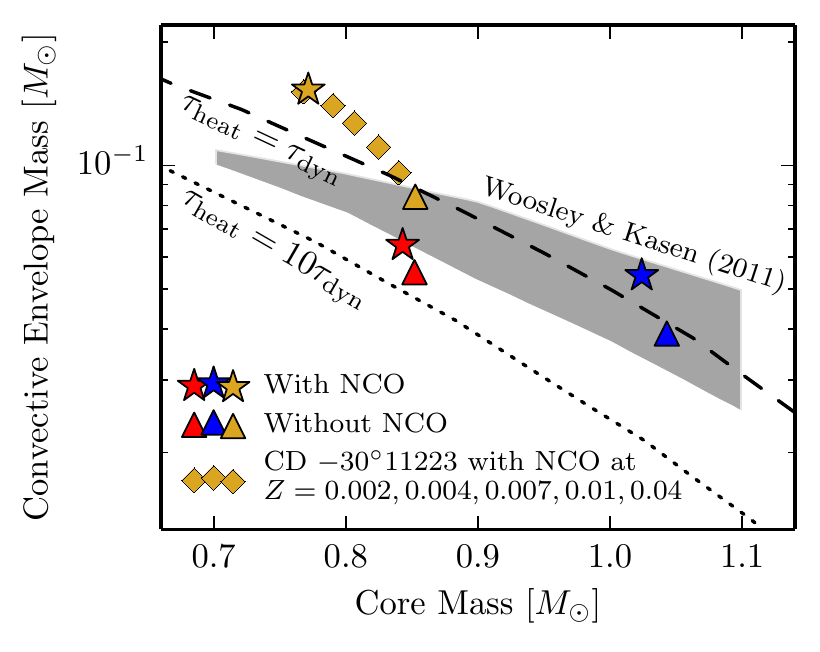}
\caption{
Masses from Table~\ref{tab:mass} showing that NCO
burning pushes convective shells toward masses that will
be more dynamical.
Red points are for the ${0.4 \, M_\odot}$~He~+ ${0.8 \, M_\odot}$~WD system, and
blue points are for the ${0.4 \, M_\odot}$~He~+ ${1.0 \, M_\odot}$~WD system.
Gold points are for CD~$-30^\circ 11223$, with diamonds used to show
models with metallicity other than solar.
The grey shaded region shows the minimum allowed envelope masses for
detonation or deflagration found by \cite{Woosley11} for a range of
core temperatures.
Note that core mass in this
figure is defined as all of the mass inside the convective shell.
}
\label{fig:dynamics}
\end{center}
\end{figure}

Following \cite{Bildsten07}, we study the possibility of dynamical burning that
may transition into a detonation by plotting the
convective shell masses from Table~\ref{tab:mass} along with lines comparing
the dynamic timescale to the heating timescale from burning in Figure~\ref{fig:dynamics}.
These timescales are defined as $\tau_{\rm dyn} = H/c_{\rm sound}$ and
$\tau_{\rm heat} = c_P T/\varepsilon_{\rm nuc}$, where $H = P/\rho g$, and $\varepsilon_{\rm nuc}$ is
dominated by $3\alpha$ burning that takes over once the thermonuclear runaway
is initiated.
These quantities are evaluated at the base of the convective burning shell.
These lines are taken from figure~5 in \cite{Shen09}.
See also figure~7 in \cite{Brooks15} for comparison, which has three points matching
the three triangular symbols on our Figure~\ref{fig:dynamics}, corresponding to
binary models that did not include NCO reactions.
For comparison, we also include the range of minimum allowed convective envelope
masses for detonation or deflagration found by \cite{Woosley11} in their extensive
grid of constant $\dot M$ models.

The solar metallicity model for CD~$-30^\circ 11223$ (gold star in Figure~\ref{fig:dynamics})
is likely to reach especially dynamical burning conditions.
At $0.1 \, \rm s$ before the most rapid evolution occurs in the \mesa\ model,
$\tau_{\rm heat}$ becomes shorter than $\tau_{\rm dyn}$.
If allowed to evolve beyond that point, peak burning reaches $\tau_{\rm heat} \approx 0.1 \tau_{\rm dyn}$,
and convective velocities reach nearly the sound speed.
We expect that a detonation should develop around this point, but hydrostatic
1D \mesa\ calculations are not reliable for evolution beyond this point.
For more detailed study of the conditions of dynamical burning and convection
leading to potential ignition of a detonation, see the recent work of \cite{Jacobs16}.
The outcomes of their 3D hydrodynamical simulations appear to be broadly consistent 
with our expectations based on Figure~\ref{fig:dynamics}, but further study is warranted.

The models for binary systems described in Sections~\ref{sec:FirstFlash} and~\ref{sec:CD-30} assumed solar metallicity for both the donor star and the WD progenitor,
but the metallicity of CD~$-30^\circ 11223$ is not known.
Varying the metallicity of the system changes the amount of time necessary for the NCO chain to deposit enough heat to initiate a runaway, and hence 
the metallicity influences both total accreted mass and convective shell mass. This trend is clear in the diamond symbols showing convective
shell masses in Figure~\ref{fig:dynamics},
which represent the binary system described in Section~\ref{sec:CD-30} modeled with NCO reactions over a range of metallicities.
For $Z \lesssim 0.02$, there is a continuous progression toward larger convective shell masses as metallicity increases for this system.
However, once $Z \gtrsim 0.02$,  there is plenty of $^{14} \rm N$ present for the NCO chain to ignite runaway burning quickly after electron captures get underway past
the threshold density, and hence higher metallicity does not significantly change the burning outcome past this point of saturation around solar metallicity.

\section{Conclusions and Future Work}

Our results show that NCO reactions play an important role in
triggering envelope ignitions for WDs accreting He at rates in the range
${1-5 \times 10^{-8} \, M_\odot/\rm yr}$. Binary systems composed of
an sdB star donating He onto a WD naturally give rise to accretion
rates that vary within this range. Because the thermal time is
comparable to the accretion time in these He envelopes, it is
necessary to model systems in a way that consistently tracks both the
full accretion rate history and the evolution of the WD in
response. Extrapolations from results using constant accretion rates
are inadequate.

Though studies at constant $\dot M$ have concluded that NCO reactions
provide only minor corrections with no qualitative differences, binary
evolution with $\dot M$ that varies over the accumulation phase shows
that NCO can be more important than previously thought.
Models for the observed system CD~$-30^\circ 11223$ illustrate the
most pronounced qualitative differences that can arise as a result of
NCO triggered ignitions during binary evolution, with a convective envelope
mass that is twice as large in the case that includes NCO reactions.
The recent discovery by \cite{Kupfer17} of another system with a tight (87 min) sdB+WD binary demonstrates that detailed binary modeling with NCO reactions will continue to remain important as more of these systems are discovered.

The dynamical nature of the most extreme flashes presented here
suggests that NCO triggered ignitions can lead to He detonations.
We have also found that metallicity is directly correlated with the
potential outcomes in these systems, with solar metallicity
progenitors providing ample fuel for the NCO chain, while lower
metallicities predictably soften its effects.
Helium shell detonations are still possible in low metallicity environments without NCO triggers,
but they appear less likely according to Figure~\ref{fig:dynamics}.
Though we have not exhaustively studied this trend, it may suggest a correlation between higher metallicity environments
and observations involving events with thick helium shell detonations.

Future exploration should include modeling of additional variables
that can impact the binary evolution of sdB+WD systems.
Models presented here assumed
that the sdB star in CD~$-30^\circ 11223$ is at the beginning of core He
burning, but this age is not
constrained. If the sdB star has a different He core burning age, the
system will make contact at a different stage of the sdB life cycle,
and the resulting accretion rate will vary accordingly. Since our
results here show that the details of the accretion rate are crucial,
variations in the $\dot M$ predicted by binary evolution may lead
to different results for the total convective shell mass ignited on the WD. 
We anticipate a potential variation of predicted outcomes for
the final fate of any given system where the sdB star age is
not known.

More work is also necessary to
strengthen our understanding of evolution beyond the onset of
dynamical burning. Understanding the transition into detonation and
the resulting effects on both the envelope and core
will be crucial for making a specific prediction
about the ultimate observable nature of these NCO triggered events.

\acknowledgements
We thank Jared Brooks for many helpful discussions on use of the
binary capabilities in \MESA\ for modeling AM CVn systems and for
providing accretion rate histories from his binary
results.
We thank Bill Wolf for developing excellent tools for scripting 
the many \MESA\ runs that went into Section~\ref{sec:constmdot},
as well as helping plot the results.
We thank Gabriel Mart{\'{\i}}nez-Pinedo for providing a machine
readable version of the $^{14}{\rm N}(e^-, \nu) {^{14}{\rm C}}$ rate
and for helpful communications regarding matrix elements.  We thank
Hendrik Schatz for helpful discussions regarding the
${^{14}{\rm C}} ( \alpha, \gamma) {^{18}{\rm O}}$ rate.  Support for
this work was provided by NASA through Hubble Fellowship grant \#
HST-HF2-51382.001-A awarded by the Space Telescope Science Institute,
which is operated by the Association of Universities for Research in
Astronomy, Inc., for NASA, under contract NAS5-26555.
This work was supported by the National Science Foundation under grants PHY 11-25915 and ACI 13-39581.
This research is funded in part by the Gordon and Betty Moore Foundation
through Grant GBMF5076 to L.B.. 

\software{
  \mesa\ \citep{Paxton11,Paxton13,Paxton15,Paxton16},
  Matplotlib \citep{Matplotlib},
  Numpy \citep{Numpy},
  MesaScript \citep{Mesascript}
}

\bibliographystyle{aasjournal}
\bibliography{evan,josiah}

\begin{thebibliography}{}
\expandafter\ifx\csname natexlab\endcsname\relax\def\natexlab#1{#1}\fi
\providecommand{\url}[1]{\href{#1}{#1}}

\bibitem[{{Ajzenberg-Selove}(1991)}]{AjzenbergSelove91}
{Ajzenberg-Selove}, F. 1991, Nuclear Physics A, 523, 1

\bibitem[{{Bildsten} {et~al.}(2007){Bildsten}, {Shen}, {Weinberg}, \&
  {Nelemans}}]{Bildsten07}
{Bildsten}, L., {Shen}, K.~J., {Weinberg}, N.~N., \& {Nelemans}, G. 2007,
  \apjl, 662, L95

\bibitem[{Brooks {et~al.}(2015)Brooks, Bildsten, Marchant, \&
  Paxton}]{Brooks15}
Brooks, J., Bildsten, L., Marchant, P., \& Paxton, B. 2015, \apj, 807, 74

\bibitem[{{Cyburt} {et~al.}(2010){Cyburt}, {Amthor}, {Ferguson}, {Meisel},
  {Smith}, {Warren}, {Heger}, {Hoffman}, {Rauscher}, {Sakharuk}, {Schatz},
  {Thielemann}, \& {Wiescher}}]{Cyburt10}
{Cyburt}, R.~H., {Amthor}, A.~M., {Ferguson}, R., {et~al.} 2010, \apjs, 189,
  240

\bibitem[{{Fuller} {et~al.}(1980){Fuller}, {Fowler}, \& {Newman}}]{Fuller80}
{Fuller}, G.~M., {Fowler}, W.~A., \& {Newman}, M.~J. 1980, \apjs, 42, 447

\bibitem[{{Fuller} {et~al.}(1985){Fuller}, {Fowler}, \& {Newman}}]{Fuller85}
---. 1985, \apj, 293, 1

\bibitem[{Geier {et~al.}(2013)Geier, Marsh, Wang, Dunlap, Barlow, Schaffenroth,
  Chen, Irrgang, Maxted, Ziegerer, Kupfer, Miszalski, Heber, Han, Shporer,
  Telting, G{\"{a}}nsicke, {\O}stensen, O’Toole, \& Napiwotzki}]{Geier13}
Geier, S., Marsh, T.~R., Wang, B., {et~al.} 2013, \aap, 554, A54

\bibitem[{Hashimoto {et~al.}(1986)Hashimoto, Nomoto, Arai, \&
  Kaminisi}]{Hashimoto86}
Hashimoto, M.~A., Nomoto, K.~I., Arai, K., \& Kaminisi, K. 1986, \apj, 307, 687

\bibitem[{Hunter(2007)}]{Matplotlib}
Hunter, J.~D. 2007, Computing In Science \& Engineering, 9, 90.
\newblock \url{https://doi.org/10.5281/zenodo.248351}

\bibitem[{Iben {et~al.}(1987)Iben, Nomoto, Tornambe, \& Tutukov}]{Iben87}
Iben, I., Nomoto, K., Tornambe, A., \& Tutukov, A.~V. 1987, \apj, 317, 717

\bibitem[{{Iliadis} {et~al.}(2010){Iliadis}, {Longland}, {Champagne}, {Coc}, \&
  {Fitzgerald}}]{Iliadis10}
{Iliadis}, C., {Longland}, R., {Champagne}, A.~E., {Coc}, A., \& {Fitzgerald},
  R. 2010, Nuclear Physics A, 841, 31

\bibitem[{{Jacobs} {et~al.}(2016){Jacobs}, {Zingale}, {Nonaka}, {Almgren}, \&
  {Bell}}]{Jacobs16}
{Jacobs}, A.~M., {Zingale}, M., {Nonaka}, A., {Almgren}, A.~S., \& {Bell},
  J.~B. 2016, \apj, 827, 84

\bibitem[{{Johnson} {et~al.}(2009){Johnson}, {Rogachev}, {Mitchell}, {Miller},
  \& {Kemper}}]{Johnson09}
{Johnson}, E.~D., {Rogachev}, G.~V., {Mitchell}, J., {Miller}, L., \& {Kemper},
  K.~W. 2009, \prc, 80, 045805

\bibitem[{{Kupfer} {et~al.}(2017){Kupfer}, {van Roestel}, {Brooks}, {Geier},
  {Marsh}, {Groot}, {Bloemen}, {Prince}, {Bellm}, {Heber}, {Bildsten},
  {Miller}, {Dyer}, {Dhillon}, {Green}, {Irawati}, {Laher}, {Littlefair},
  {Shupe}, {Steidel}, {Rattansoon}, \& {Pettini}}]{Kupfer17}
{Kupfer}, T., {van Roestel}, J., {Brooks}, J., {et~al.} 2017, \apj, 835, 131

\bibitem[{{Mart{\'{\i}}nez-Rodr{\'{\i}}guez}
  {et~al.}(2016){Mart{\'{\i}}nez-Rodr{\'{\i}}guez}, {Piro}, {Schwab}, \&
  {Badenes}}]{MartinezRodriguez2016}
{Mart{\'{\i}}nez-Rodr{\'{\i}}guez}, H., {Piro}, A.~L., {Schwab}, J., \&
  {Badenes}, C. 2016, \apj, 825, 57

\bibitem[{{Nelemans} {et~al.}(2004){Nelemans}, {Yungelson}, \& {Portegies
  Zwart}}]{Nelemans04}
{Nelemans}, G., {Yungelson}, L.~R., \& {Portegies Zwart}, S.~F. 2004, \mnras,
  349, 181

\bibitem[{{Neunteufel} {et~al.}(2016){Neunteufel}, {Yoon}, \&
  {Langer}}]{Neunteufel16}
{Neunteufel}, P., {Yoon}, S.-C., \& {Langer}, N. 2016, \aap, 589, A43

\bibitem[{{Nomoto}(1982)}]{Nomoto82}
{Nomoto}, K. 1982, \apj, 253, 798

\bibitem[{{Paxton} {et~al.}(2011){Paxton}, {Bildsten}, {Dotter}, {Herwig},
  {Lesaffre}, \& {Timmes}}]{Paxton11}
{Paxton}, B., {Bildsten}, L., {Dotter}, A., {et~al.} 2011, \apjs, 192, 3

\bibitem[{{Paxton} {et~al.}(2013){Paxton}, {Cantiello}, {Arras}, {Bildsten},
  {Brown}, {Dotter}, {Mankovich}, {Montgomery}, {Stello}, {Timmes}, \&
  {Townsend}}]{Paxton13}
{Paxton}, B., {Cantiello}, M., {Arras}, P., {et~al.} 2013, \apjs, 208, 4

\bibitem[{{Paxton} {et~al.}(2015){Paxton}, Marchant, Schwab, Bauer, Bildsten,
  Cantiello, Dessart, Farmer, Hu, Langer, Townsend, Townsley, \&
  Timmes}]{Paxton15}
{Paxton}, B., Marchant, P., Schwab, J., {et~al.} 2015, \apjs, 220, 15

\bibitem[{{Paxton} {et~al.}(2016){Paxton}, {Marchant}, {Schwab}, {Bauer},
  {Bildsten}, {Cantiello}, {Dessart}, {Farmer}, {Hu}, {Langer}, {Townsend},
  {Townsley}, \& {Timmes}}]{Paxton16}
{Paxton}, B., {Marchant}, P., {Schwab}, J., {et~al.} 2016, \apjs, 223, 18

\bibitem[{Piersanti {et~al.}(2001)Piersanti, Cassisi, \&
  Tornamb\'{e}}]{Piersanti01}
Piersanti, L., Cassisi, S., \& Tornamb\'{e}, A. 2001, \apj, 558, 916

\bibitem[{{Piro}(2015)}]{Piro15}
{Piro}, A.~L. 2015, \apj, 801, 137

\bibitem[{{Potekhin} {et~al.}(2009){Potekhin}, {Chabrier}, \&
  {Rogers}}]{Potekhin09a}
{Potekhin}, A.~Y., {Chabrier}, G., \& {Rogers}, F.~J. 2009, \pre, 79, 016411

\bibitem[{Schwab {et~al.}(2015)Schwab, Quataert, \& Bildsten}]{Schwab15}
Schwab, J., Quataert, E., \& Bildsten, L. 2015, \mnras, 453, 1910

\bibitem[{{Schwab} {et~al.}(2016){Schwab}, {Quataert}, \&
  {Bildsten}}]{Schwab16}
{Schwab}, J., {Quataert}, E., \& {Bildsten}, L. 2016, \mnras, 458, 3613

\bibitem[{Shen \& Bildsten(2009)}]{Shen09}
Shen, K.~J., \& Bildsten, L. 2009, \apj, 699, 1365

\bibitem[{Shen \& Bildsten(2014)}]{Shen14}
---. 2014, \apj, 785, 61

\bibitem[{van~der Walt {et~al.}(2011)van~der Walt, Colbert, \&
  Varoquaux}]{Numpy}
van~der Walt, S., Colbert, S.~C., \& Varoquaux, G. 2011, Computing in Science
  \& Engineering, 13, 22.
\newblock \url{https://pypi.python.org/pypi/numpy/1.12.1}

\bibitem[{Wolf {et~al.}(2017)Wolf, Bauer, \& Schwab}]{Mesascript}
Wolf, B., Bauer, E.~B., \& Schwab, J. 2017, wmwolf/MesaScript: A DSL for
  Writing MESA Inlists, v1.0.2,  Zenodo, doi:10.5281/zenodo.826954.
\newblock \url{https://doi.org/10.5281/zenodo.826954}

\bibitem[{Woosley \& Kasen(2011)}]{Woosley11}
Woosley, S.~E., \& Kasen, D. 2011, \apj, 734, 38

\bibitem[{Woosley \& Weaver(1994)}]{Woosley94}
Woosley, S.~E., \& Weaver, T.~A. 1994, \apj, 423, 371

\bibitem[{{Yoon} \& {Langer}(2004{\natexlab{a}})}]{Yoon04a}
{Yoon}, S.-C., \& {Langer}, N. 2004{\natexlab{a}}, \aap, 419, 645

\bibitem[{{Yoon} \& {Langer}(2004{\natexlab{b}})}]{Yoon04b}
---. 2004{\natexlab{b}}, \aap, 419, 623

\bibitem[{{Yoon} {et~al.}(2004){Yoon}, {Langer}, \& {Scheithauer}}]{Yoon04c}
{Yoon}, S.-C., {Langer}, N., \& {Scheithauer}, S. 2004, \aap, 425, 217

\end{thebibliography}

\end{document}